\def\bd{\begin{displaymath}}\def\ed{\end{displaymath}}
\def\be{\begin{equation}}\def\ee{\end{equation}}
\def\bea{\begin{eqnarray}}\def\eea{\end{eqnarray}}
\def\lb{\label}

\def\k{\kappa}
\def\x{\xi}

\def\D{\Delta}

\def\de{\partial}
\def\id{\equiv}

\def\ex{{\rm e}}

\def\poi{Poincar\'e }

\def\PL#1{Phys.\ Lett.\ {\bf#1}}
\def\PRL#1{Phys.\ Rev.\ Lett.\ {\bf#1}}
\def\PR#1{Phys.\ Rev.\ {\bf#1}}

 \def\IJMP#1{Int.\ J. Mod.\ Phys.\ {\bf #1}}
\def\MPL#1{Mod.\ Phys.\ Lett.\ {\bf #1}} 

\def\AoP#1{Ann.\ Phys.\ {\bf#1}}
\def\grq#1{{\tt gr-qc/\-#1}}\def\hep#1{{\tt hep-th/\-#1}}

\def\den{\left(1-{E\over\k}\right)}
\def\epe{\ex^{{E\over\k}}}
\def\eme{\ex^{-{E\over\k}}}
\def\mk{{m_0\over\k}}\def\mks{\mbox{\small$\mk$}}

\documentclass[12pt]{article}
\begin{document}

\begin{titlepage}
\vspace{.3cm}
\begin{center}
\renewcommand{\thefootnote}{\fnsymbol{footnote}}
{\Large \bf On the definition of velocity in theories with two
observer-independent scales}
\vfill
{\large \bf {S.~Mignemi\footnote{email: smignemi@unica.it}}}\\
\renewcommand{\thefootnote}{\arabic{footnote}}
\setcounter{footnote}{0}
\vfill
{\small
  Dipartimento di Matematica, Universit\`a di Cagliari,\\
Viale Merello 92, 09123 Cagliari, Italy\\
\vspace*{0.4cm}
 INFN, Sezione di Cagliari\\
}
\end{center}
\vfill
\centerline{\bf Abstract}
\vfill

We argue that a consistent definition of the velocity of a
particle in generalizations of special relativity with two
observer-independent scales
should be independent from the mass of the particle. This
request rules out the definition $v_i=\de p_0/\de p_i$, but allows
for other definitions proposed in the literature.

\vfill
\end{titlepage}

\section{Introduction}
Recently, following a suggestion of Amelino-Camelia \cite{AC},
large interest has been devoted to modifications of special
relativity admitting two observer-independent
fundamental scales, the speed of light and the Planck energy.
These models aim to describe the dynamics of particles up to the
Planck region, where the structure of spacetime may change due
to quantum gravity effects.

The existence of two observer-independent scales is implemented
in the theory through a nonlinear action of the Lorentz group
on momentum space, whose main consequence is a deformation of the
dispersion relations of special relativity, which are recovered
only in the low-energy limit.
Of course, it is possible to construct several different models
obeying these postulates.
Historically, the first example  was given by the quantum \poi
algebra of Lukierski, Nowicki and Ruegg (LNR) \cite{LNR}.
More recently, an algebraically simpler model has been introduced
by Magueijo and Smolin (MS) \cite{MS}.

The identification of physical quantities in these models may lead
to problems, since the full range of validity of the theory is not
accessible to experiments, and different definitions may lead to
the same low-energy limit. In absence of experimental tests,
one has to resort to requirements of consistency.

A debated problem is for example the correct definition of
the velocity of a particle. Several different
proposal have been advanced in the literature [1-7].
Apparently, the most natural proposal is to define the velocity
like in Hamiltonian mechanics as $v_H\id{\de E\over\de p}$ \cite{AC}.
However, this proposal gives rise to complicated addition laws
for velocity  and, as we shall see, does not seem to be
fully consistent. Other proposal have been advanced [4-6],
which instead predict the classical addition law of special
relativity.

In this note, we suggest that a good definition of velocity
should be such that any particles having the same velocity in a
reference frame must have the same velocity in any other frame.
In particular, the Lorentz transformation between the rest frame
of a particle and a frame where it moves with velocity $v$,
should depend only on $v$ and not on the mass of the particle.
This request singles out the definitions of refs.\ [4-6].
In particular, the hamiltonian definition $v=v_H$ is ruled out.
Similar conclusions have been reached in ref. \cite{KM} starting
from a totally different point of view.

In the following, we consider for simplicity of notation a
two-dimensional spacetime (generalization to four dimensions is
trivial).
We use $(+,-)$ signature and denote with $(E,p)$ the components
of the 2-momentum $p_a$, $a=0,1$.

\section{The MS model}
We start the discussion from the MS model \cite{MS}. In this case,
a boost of rapidity parameter $\x$ is assumed to transform the
2-momentum $(E_0,p_0)$ into $(E,p)$, where
\be\lb{MSt}
E={E_0\cosh\x-p_0\sinh\x\over\D},\qquad
p={p_0\cosh\x+E_0\sinh\x\over\D},
\ee
with $\D=1+(E_0(\cosh\x-1)-p_0\sinh\x)/\k$.

The Casimir mass $m$, defined as
\be\lb{MSc}
m^2={E^2-p^2\over\den^2},
\ee
is invariant under the transformations (\ref{MSt}), and is related
to the rest energy $m_0$ of the particle by
\be
m_0={m\over1+{m\over\k}}.
\ee

Consider now a particle of Casimir mass $m$ at rest in an inertial
frame. Its 2-momentum is given by $(m_0,0)$.
From (\ref{MSt}), we can derive the energy $E$ and the momentum $p$
of the particle in
a frame related to the first by a boost of parameter $\x$:
\be\lb{MSe}
E={m\cosh\x\over1+{m\over\k}\cosh\x},\qquad
p={m\sinh\x\over1+{m\over\k}\cosh\x}.
\ee
One can then express $\cosh\x$ in terms of $E$
and $m$ \cite{AK}:
\be
\cosh\x={E\over m\den},
\ee
to be compared with the classical result, $\cosh\x=E/m$.

However, we are interested in the relation between the rapidity
$\x$ and the velocity $v$ of the particle in the moving frame.
If the velocity of the particle is defined by the relation
$v_H={\de E\over\de p}$, one can derive its expression
differentiating the mass-shell constraint (\ref{MSc}),
$p^2=E^2-m^2\den^2$.
One has
\be
v_H={p\over E+{m^2\over\k}\den},
\ee
and hence, from (\ref{MSe}),
\be\lb{MSv}
v_H={\sinh\x\over\cosh\x+{m\over\k}}.
\ee
Inverting (\ref{MSv}) one obtains the rapidity parameter in terms
of the velocity:
\be
\cosh\x={{m\over\k}v_H^2+\sqrt{1-\left(1-{m^2\over\k^2}\right)v_H^2}
\over1-v_H^2}.
\ee
Hence, the rapidity parameter of the boost that relates the rest
frame to the frame where the particle has velocity $v_H$ depends on
the mass of the particle and particles of different masses at rest
in one frame would have different velocities in another frame.
Moreover, there is no definite relation between $\x$ and the
velocity of the moving frame.

This problem is not present if one defines the velocity as in
\cite{Gr} (but this definition was already implicit in \cite{MS}),
\be
v_G={p\over E}.
\ee
In fact, in view of eq.\ (\ref{MSe}),
\be
v_G=\tanh\x,
\ee
from which follows the classical relation of special relativity,
\be
\cosh\x={1\over\sqrt{1-v_G^2}},
\ee
which is of course independent from the particle mass.

\section{The LNR model}
We pass now to consider the LNR model. In this case, the
transformation laws of the 2-momentum are \cite {BAK}
\be\lb{LNt}\
\epe=\D\,{\ex^{E_0\over\k}},\qquad
p={p_0\cosh\x+\k\left(1-\cosh\mk\,\eme\right)\sinh\x\over\D},
\ee
where
\be
\D=\cosh\mks\,\eme+\left(1-\cosh\mks\,\eme\right)\cosh\x+{p\over\k}
\sinh\x,
\ee
and $m_0$ is the rest energy of the particle, which in terms
of the invariant Casimir mass $m$, defined as
\be\lb{LNc}
m^2=\k^2\left(\epe-2+\eme\right)-p^2\epe,
\ee
is given by
\be
\cosh\mk=1+{m^2\over2\k^2}.
\ee

Consider now a particle at rest in an inertial
frame, and derive from (\ref{LNt}) its energy $E$ and
momentum $p$ in a frame related to the first by a boost
of rapidity parameter $\x$:
\bea
&&\epe={\cosh\mks+\sinh\mks\cosh\x},\lb{LNe}\\
&&p={\sinh\mk\sinh\x\over\cosh\mk+\sinh\mk\cosh\x}.\lb{LNp}
\eea
From (\ref{LNe}), one can express $\cosh\x$ in terms of $E$
and $m_0$ \cite{AK,AA}:
\be
\cosh\x={\epe-\cosh\mk\over\sinh\mk}.
\ee

One can derive now the relation between the rapidity parameter
$\x$ and the velocity $v$ of the particle. If the velocity is
defined by the relation
$v_H={\de E\over\de p}$, one can obtain its expression
differentiating the mass-shell constraint (\ref{LNc}),
$p^2=\k^2\left(1-\eme\right)^2-m^2\eme$.
One has
\be
v_H={p\,\epe\over\k\left(\cosh\mk-\eme\right)},
\ee
and hence, from (\ref{LNe}) and (\ref{LNp})
\be\lb{LNv}
v_H={\cosh\mks+\sinh\mks\cosh\x\over\sinh\mk+\cosh\mk\cosh\x}\,\sinh\x.
\ee
Of course, inverting (\ref{LNv}) one obtains that the rapidity parameter
depends both on the velocity and the mass of the particle, leading to the
same problems as with the MS model.

Also in this case these problems can be avoided adopting a
different definition for the velocity of a particle.
Namely, one can adopt the right velocity introduced in \cite{THM,LN},
\be
v_R\id{v_H\over 1+{p\,v_H\over\k}}={p\,\epe\over\k\left(\epe-\cosh\mk\right)}.
\ee
After substituting (\ref{LNe}) and (\ref{LNp}), in fact, one obtains
\be
v_R=\tanh\x,\qquad\cosh\x={1\over\sqrt{1-v_R^2}},
\ee
which are again the classical relations, independent from the particle mass..

\section{Conclusions}
We have shown that in order to obtain a definite relation between
the relative velocity of two reference frames and the rapidity of
a boost relating them, one cannot define the velocity of a particle
of energy $E$ and momentum $p$ as $v={\de E\over\de p}$. A suitable
definition seems to be model-dependent (see however \cite{KM}),
but in the known cases always
satisfies the special relativistic relations, and in particular the
addition law of velocity \cite{Gr,LN}. Although this is of course a
sufficient condition for avoiding the problems discussed in this
paper, it does not appear to be also necessary in principle.

It is also interesting to notice that the correct definition of velocity
implies that the velocity of a massless particle is always equal
to $c$. In our opinion, this is a very basic prediction and we see no
reason for introducing a variable speed of light.
First of all, this would be at odds with the spirit of the model,
which is based on the existence of two invariant fundamental scales.
A more serious problem is that a variable speed of light would
destroy the causal structure of special relativity, leading to great
difficulties with the physical interpretation.

Of course, it would be useful to build a suitable hamiltonian
formalism that predicts the correct velocities. This seems to be
possible only if one uses deformed Poisson brackets and
noncommuting spacetime coordinates \cite{KG,LN,Gr}.

\end{document}